\documentclass[twoside]{article}
\usepackage{fleqn,espcrc2}

\usepackage{graphicx}
\usepackage[figuresright]{rotating}

\title{Quarkonium production and NRQCD matrix elements}

\author{Adam K.\ Leibovich{\\Department of Physics, 
Carnegie Mellon University,\\
 Pittsburgh, Pennsylvania 15213, USA}}
       
\begin{document}

\begin{abstract}
Most recent calculations of quarkonium production are based on the
NRQCD factorization formalism.  This formalism is reviewed.  To make
predictions about specific cross section, universal NRQCD matrix
elements need to be extracted from experiments.  Extractions from
different experimental situations are compared, with some emphasis on
the extraction from LEP.
\vspace{1pc}
\end{abstract}

\maketitle

\section{INTRODUCTION}
Quarkonium is an interesting system since the mass of the heavy quark
is much larger than the QCD scale,
\begin{equation}
m_Q \gg \Lambda_{\rm QCD}.
\end{equation}
One would hope, therefore, that it would be possible to calculate
reliably in perturbation theory.  If that were the end of the story,
things would be simple.  The only difficulty, which makes quarkonium
very interesting, is that there is another natural small scale in the
problem, the relative velocity of the heavy quarks, $v$.  For
charmonium $v_c^2 \sim 1/4$, while for bottomonium $v_b^2 \sim 1/10$.

Since we have a multi-scale problem, with the hierarchy
\begin{equation}
m_Q \gg m_Q v \gg m_Q v^2 \sim \Lambda_{\rm QCD},
\end{equation}
it is useful to use an effective field theory.  The effective field
theory appropriate for a system with two heavy quarks is
Nonrelativistic QCD (NRQCD) \cite{BBL,recent}.  As the name implies,
there is a relativistic expansion in $v$, as well as the usual
perturbative expansion in $\alpha_s$.

Before the advent of NRQCD, quarkonium production and decay was
calculated using the so-called ``Color-Singlet Model'' (CSM).  In this
model, the $\bar QQ$ pair is treated in a color-singlet state, in the
$v\to0$ limit.  It was discovered, however, that the CSM did not
describe quarkonium production very well in certain circumstances.
For example, the CSM prediction for $\psi'$ production at the Tevatron
is off by about a factor of 30!  As we will see, NRQCD can easily explain
the discrepancy in this situation \cite{BF}.

Whether NRQCD is the correct effective field theory to be used for
quarkonium is an interesting and important question.  There are some
potential problems comparing NRQCD predictions to data, in particular
the polarization.  NRQCD make a definite prediction that at large
transverse momenta, $\psi'$ should be nearly 100\% polarized
\cite{CW}.  However, it appears that the data is not following that
trend \cite{Affolder}.  So it may be that the $\alpha_s$ and $v$
expansions do not converge well for charmonium.  This is not to say
that NRQCD is incorrect.  NRQCD is a valid effective field theory, in
that as the mass $m_Q$ goes to infinity, the theory correctly
reproduces full QCD.  It may be that $v_c$ is too large, or $m_Q
v_c^2$ too small, for NRQCD to be useful in describing charmonium
production.  One could still hope that it will work for bottomonium.

How is it possible to test NRQCD?  There is the polarization discussed
above \cite{Kramer}.  Another way is to compare NRQCD matrix elements
extracted from different experiments.

\section{FACTORIZATION FORMALISM}
The physical picture of production in NRQCD begins with a hard
scattering, in which a $\bar QQ$ pair are produced with any spin,
angular and color quantum numbers.  This process can be calculated in
perturbation theory.  Then the $\bar QQ$ evolve into the final state
quarkonium $H$ in some non-perturbative fashion.  This is encoded in
the NRQCD matrix elements, which must be extracted from experiment.

The NRQCD factorization formula is the mathematical realization of the
above picture.  A general production process for
$i+j\to H+X$ can be written as
\begin{equation}
d\sigma = \sum_n 
   d\hat{\sigma}(ij\to Q\bar{Q}[n]+X) \langle O^H(n)\rangle,
\end{equation}
where $d\hat{\sigma}(ij\to Q\bar{Q}[n]+X)$ is the cross section for
producing $Q\bar{Q}$ in state $n$ by scattering $i$ and $j$,
calculable in a perturbative expansion in $\alpha_s$ and perhaps
convoluted with parton distribution functions (PDFs).  The long
distance matrix elements encode the hadronization of the heavy quarks
in state $n$ into the final quarkonium state $H$.  The matrix elements
can be written as
\begin{eqnarray}
\langle O^H(n)\rangle &=& 
   \langle0|\psi^\dag\Gamma^n\chi|\sum_X H+X\rangle \nonumber\\
&& \times \langle\sum_X H+X|\chi^\dag{\Gamma^n}'\psi|0\rangle
\end{eqnarray}
The $\Gamma^n$ contains Dirac matrices, color matrices and
derivatives.  The NRQCD matrix elements scale with a definite power of
$v$, determined by $\Gamma^n$, allowing the truncation of the $v$
expansion.  Again, we expect the velocity expansion will work better
in the bottom sector than the charm sector.  The factorization
described above still needs to be put on firmer ground.  We should
keep in mind that when we use the NRQCD factorization formalism to
predict production rates, we are not only testing the $\alpha_s$ and
$v$ expansions, but also the factorization formalism.

The scaling is determined by looking at what perturbations are
necessary to give a non-vanishing result for the time ordered product.
To give non-zero overlap, multipole moment interactions may need to be
inserted.  For example, the matrix element $\langle
O_8^{J/\psi}(^3S_1)\rangle$ scales as $v^4$ compared to the $\langle
O_1^{J/\psi}(^3S_1)\rangle$, since we need two $E1$ insertions in the
amplitude.  The first insertion changes $L$ and neutralizes the color.
The second changes $L$ back to an $S$ wave.  Each insertion cost a
factor of $v$ in the amplitude, or a total of $v^4$ in the rate.  The
scalings of the most important matrix elements for $\psi$ production
are
\begin{eqnarray}
\langle O_1^{\psi}(^3S_1)\rangle &\sim& v^0, \\
\langle O_8^{\psi}(^3S_1)\rangle &\sim& v^4, \\
\langle O_8^{\psi}(^3P_J)\rangle &\sim& v^4, \\
\langle O_8^{\psi}(^1S_0)\rangle &\sim& v^4.
\end{eqnarray}

We can relate the color-singlet matrix elements to the wave function at
the origin, or derivatives of the wave function at the origin, using
the vacuum saturation approximation
\begin{equation}
\langle O^H(n)\rangle \approx 
  \langle0|\psi^\dag\Gamma^n\chi|H\rangle
  \langle H|\chi^\dag\Gamma^n\psi|0\rangle,
\end{equation}
dropping the sum over $X$.  This allows us to use the lattice, data,
or models to obtain the color-singlet matrix elements.  It also means
that the color-singlet model has been incorporated into the NRQCD
factorization formalism.  

A quick aside.  Many people say the NRQCD model, or the
``color-octet'' model.  However, NRQCD is not a model.  The CSM, on
the other hand, is a model, since there is no limit in which it will
reproduce full QCD.

\section{QUARKONIUM AT THE TEVATRON}
As previously mentioned, the CSM fails to describe the data at the
Tevatron.  We will now discuss how NRQCD improves the situation.  We
will concentrate on $J/\psi$ production.  A similar analysis can be
done for $\Upsilon$ production.  See Refs.~\cite{DSL} for recent analyses.

\begin{table*}[htb]
\caption{Table for $M_k^{J/\psi}$ in units of $10^{-2}{\rm\ GeV}^3$
extracted from the Tevatron data.  The second set of errors, when
present, are due to scale variation.  The first set are statistical.
$k$ varies between 3 and 3.5.}
\label{tab:Mk}
\begin{tabular}{@{}lccccccc}
\hline
Ref. & MRSD0 & MRS(R2) & CTEQ2L & CTEQ4L & GRVLO & GRVHO \\
\hline
\cite{CL1} & $6.6 \pm 1.5$ & - & - & - & - & - \\
\cite{BK}  & - & $10.09 \pm 2.07^{+2.79}_{-1.26}$  & - &  
   $4.38 \pm 1.15^{+1.05}_{-0.74}$ & 
   $3.90 \pm 1.14^{+1.46}_{-1.07}$ & - \\
\cite{CSL} & $1.32 \pm 0.21$ & - & $1.44 \pm 0.21$ & - & - &
   $0.60 \pm 0.21$ \\
\cite{SL} & $1.32 \pm 0.21$ & - & $1.32 \pm 0.21$ & - & - &
   $0.45 \pm 0.09$ \\
\cite{KK} & - & - & - & $6.52 \pm 0.67 $ & - & - \\
\hline
\end{tabular}
\end{table*}

At large transverse momentum, the leading order NRQCD prediction in
$\alpha_s$ and $v$ is $O(\alpha_s^3v^0)$.  This is the same as the
leading order CSM prediction, which grossly underestimates the rate.
Even the shape of spectrum is wrong, with the theory dropping as
$1/p_\perp^8$ compared to the data which falls as $1/p_\perp^4$.

We can go to higher order in the perturbative expansion:
$O(\alpha_s^5 v^0)$.  In this case, fragmentation processes occur,
which, due to some propagators being close to on-shell, can have
large contributions.  These fragmentation diagrams dominate over the
lower order graphs at large $p_\perp$, with the theory now falling as
$1/p_\perp^4$, similar to the data.  However, the normalization of the
theory is still far below the experiment.  

This is as far as the CSM can go.  To improve the prediction, NRQCD is
needed \cite{BF}.  Instead of just including higher order in
$\alpha_s$, we can also include higher order in $v$:
$O(\alpha_s^3v^4)$.  Now there are color-octet fragmentation processes
proportional to the matrix element $\langle O_8^\psi(^3S_1)\rangle$,
which also scales as $1/p_\perp^4$.  Of course, the value of the
matrix element is unknown, so the fact that we can get a good fit
really means that the extracted matrix element is not abnormally large
or small.

The fragmentation contribution can be approximated as
\begin{equation}
d\sigma(p\bar p\to H) = \sum_i\int dz\,d\sigma_i\,D_{i\to H},
\end{equation}
where $d\sigma_i$ is the cross section to produce an on-shell parton
$i$, $D_{i\to H}$ is the fragmentation function for parton $i$ to
produce quarkonium state $H$ with momentum fraction $z$, and the sum
is over partons.  The advantage of writing it in this way is that it
is easy to sum large logs of $\log p_\perp/m_Q$ using Altarelli-Parisi
(AP) evolution.

At lower $p_\perp$, fragmentation no longer dominates, and it is
necessary to include all non-fragmentation diagrams as well.
Contributions due to other octet matrix elements, $\langle
O_8^\psi(^1S_0)\rangle$ and $\langle O_8^\psi(^3P_J)\rangle$ are now
important.  These channels both fall as $1/p_\perp^6$, making it
impossible to determine them independently.  Instead, a linear
combination of the two
\begin{equation}
M_k^H = \langle O_8^\psi(^1S_0)\rangle 
   + \frac{k}{m_Q^2}\langle O_8^\psi(^3P_0)\rangle
\end{equation}
is usually extracted.  

At small $p_\perp$, higher twist contributions become important.
These contributions scale as powers of
$\Lambda/\sqrt{p_\perp^2+m_Q^2v^n}$, where the value of $n$ is not
known.  Also at small $p_\perp$, the rates tend to diverge, due to
soft gluon effects.  To handle these divergences, the low $p_\perp$
region can just be ignored or modeled by including initial parton
showering/intrinsic $k_\perp$.

\begin{table*}[htb]
\caption{Table for $\langle O_8^{J/\psi}(^3S_1)\rangle$ in units of
$10^{-3}{\rm\ GeV}^3$ extracted from the Tevatron data. The second set
of errors, when present, are due to scale variation.  The first set
are statistical.}
\label{tab:oct}
\begin{tabular}{@{}lcccccccc}
\hline
Ref. & MRSD0 & MRS(R2) & CTEQ2L & CTEQ4L & GRVLO & GRVHO \\
\hline
\cite{CL1} & $6.6 \pm 2.1$ & - & - & - & - & - \\
\cite{BK}  & - & $14.0 \pm 2.2^{+13.5}_{-7.9}$  & - &  
   $10.6 \pm 1.4^{+10.5}_{-5.9}$ & 
   $11.2 \pm 1.4^{+9.9}_{-5.6}$ & - \\
\cite{CSL} & $2.1 \pm 0.5$ & - & $3.3 \pm 0.5$ & - & - &
   $3.4 \pm 0.4$ \\
\cite{SL} & $6.8 \pm 1.6$ & - & $9.6 \pm 1.5$ & - & - &
   $9.2 \pm 1.1$ \\
\cite{KK} & - & - & - & $3.94 \pm 0.63 $ & - & - \\
\hline
\end{tabular}
\end{table*}

In Table \ref{tab:Mk}, different extractions of $M_k^{J/\psi}$ are
collected.  The second set of errors, when present, are due to scale
variation.  The other errors are statistical.  As can be seen, there
is large uncertainty due to the choice of PDF.  At this point, the
best we can hope to say is that this combination of matrix elements is
on the order of
\begin{equation}
M_3^{J/\psi} \sim {\rm few\ }10^{-2}{\rm\ GeV}^3.
\end{equation}

Table \ref{tab:oct} contains different extractions of $\langle
O_8^{J/\psi}(^3S_1)\rangle$.  Again the second set of errors, when
present, are due to scale variation, while the other errors are
statistical.  Here the largest error is not due to the PDFs, but due
to the scale variation.  This is an indication that higher order
perturbative corrections may be very large.  Again we only know the
matrix element is on the order of
\begin{equation}
\langle O_8^{J/\psi}(^3S_1)\rangle \sim {\rm few\ }10^{-3}{\rm\ GeV}^3.
\end{equation}

\section{LEP EXTRACTION}
We would like to do better than the order of magnitude extractions
discussed above at the Tevatron.  Since the dominant errors are due
to scale or PDF variation, it seems difficult make improvements at the
Tevatron.  We would also like to test the formalism, by comparing the
matrix elements extracted from different experiments.

The natural place to look for cleaner extractions would be at an
$e^+e^-$ machine, since there is no PDF nor initial state
gluon radiation to worry about.  The prediction at CLEO
does not have a large dependence on the color-octet matrix elements
away from the endpoint.  Therefore LEP is the natural choice.  

Formally there are two leading order contributions in the $\alpha_s$
and $v$ expansion, both in the singlet channel, of order $O(\alpha_s^2
v^0)$.  The contribution from gluon radiation in the singlet channel
$Z \to \psi g g$ is suppressed by powers of $M_\psi^2/E_\psi^2$
\cite{KS}. The color-singlet charm quark fragmentation process $Z
\to\psi c {\overline c}$ \cite{BCK}, which has no power suppression,
dominates over non-fragmentation processes for large $E_\psi$.  Light
quark octet fragmentation (in which the mother parton does not combine
to form part of the bound state) is naively of order $\alpha_s^2 v^4$,
down by $v^4 \sim 1/10$ compared to charm fragmentation.  However,
this channel is enhanced due to the presence of large logs and a
numerical factor of five due to the number of possible quarks that
initiate the process \cite{CKY}.  The same logs that enhance the octet
channel also put the convergence of the perturbative expansion into
question.

The tree-level calculation of the differential cross section in the
color octet production channel \cite{CKY} scales as $z\to1$ as
\begin{equation}
\frac{d\Gamma}{dz} \sim \alpha_s^2 
   \frac{\log(M_Z^2/M^2_\psi)}{z}\langle O_8^{J/\psi}(^3S_1)\rangle, 
\end{equation} 
leading to large double logs in the total rate. Since $\alpha_s
\log(M_Z^2/M^2_\psi) \approx 1.5$, we should treat $\alpha_s
\log(M_Z^2/M^2_\psi)$ as order one and resum all powers of the large
logarithm. With this counting, the octet channel is $O(\alpha_s^0
v^4)$, on par with the singlet fragmentation contribution.  More
practically, the tree-level calculation has a factor of two
uncertainty associated with the scale at which $\alpha_s$ is
evaluated, since $\alpha_s(M_\psi)/\alpha_s(M_Z) \approx 2$ (this is
just a restatement that there is a large logarithm).  The resummation
of the leading logarithms reduces this uncertainty, so the
resummation procedure is essential from both a practical and a formal
standpoint.

To resum the logs, we write the rate in the fragmentation limit as
\begin{eqnarray}
\frac{d\Gamma}{dz} &=& \int_z^1 \frac{dy}y \left[2 C_q(\mu^2,y) 
  D_q(\mu^2,z/y)\right.  \nonumber\\
  &&\phantom{\int_z^1 \frac{dy}y}\left.+\,C_g(\mu^2,y) D_g(\mu^2,z/y)\right].
\end{eqnarray}
We can now use AP to sum the logs of $M_Z^2/M^2_\psi$.

However, summing the above mentioned logs will only yield the correct
leading order differential rate if $z$ is sufficiently
large.\footnote{At very large $z\sim1$ we do not get the correct
result since we have neglected the shape function which enters near
the endpoint.  This shape function has been argued to be the solution
of the so called ``Hera anomaly''.  It is not important here since
the rate is very small in this region.}  When $z$ is parametrically
small, terms of the form $\alpha_s\log(z)/z$ become just as
important. Furthermore, these logs will also contribute double logs to
the total rate given that the lower limit on $z$ is $2 M_\psi/M_Z$.
This second type of log, due to soft gluon emission, is resummed using
a formalism familiar from discussions of jet multiplicities
\cite{rus}, where the color-coherence of the soft gluon emission
is very important.

It is possible to do both of these resummations \cite{BLR}, see
Fig.~1.  The rate depends on the linear combination
\begin{eqnarray}
\sum_m \langle O_8^{H(m)}(^3S_1) \rangle 
   \times {\rm Br}[H(m) \to J/\psi X] \nonumber \\
= (1.9\,\pm\,0.5_{stat}\,\pm\,1.0_{theory})\times 10^{-2}{\rm\ GeV}^3
\end{eqnarray}
since the data includes feeddown from higher charmonium states.  It
turns out that this leads to an extraction with much smaller
theoretical uncertainty than the analogous value from the Tevatron
$1.4 \pm 0.2_{stat} \pm 1.4_{theory}$.  For the Tevatron value, the
extraction from Ref.~\cite{CL1} was used for the central value.
\begin{figure}[th]
\label{fig:lep}
\includegraphics[scale=0.38]{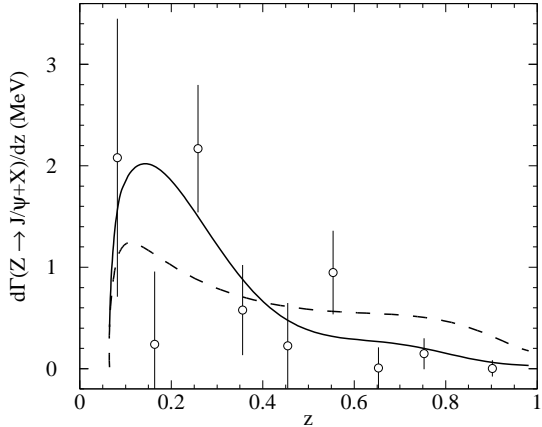}
\caption{Differential rate $d\Gamma/dz$ as a function of
$z=2E_\psi/M_Z$ vs data.  The dashed line is the sum of the tree-level
octet and singlet results and the solid line is the octet plus the
singlet resummation.  From Ref.~\cite{BLR}}
\end{figure}

It is difficult to tell from Fig.~1 if there is much
improvement in the prediction from this resummation.  To make the
comparison more quantitative, it is useful to take the ratio of the
first moment to the zeroth moment
\begin{equation}
\frac{1}{\Gamma(Z\to J/\psi X)}
   \int dz\,z\,\frac{d\Gamma}{dz}(Z\to J/\psi X)
\end{equation}
which for the resummed rate is $0.30$.  The tree level rate gives
$\sim 0.5$.  A very rough estimate of this quantity obtained from the
data \cite{ALEPH} suggests a value of $0.26 \pm 0.10$.  This is in
sharp contrast to the color singlet prediction.  The tree-level color
singlet decay rate predicts the ratio of the first moment over the
zeroth moment to be $0.62$.  Resummation softens the color singlet
decay rate, but the ratio is still too large, $0.47$.  The ratio is
independent of the color singlet matrix element.  A rigorous
extraction of the first moment by the experimental groups could
provide an extremely clean, quantitative test of the NRQCD approach.

\section{OTHER EXTRACTIONS}
\subsection{HERA}
The NRQCD prediction does not seem to fit the data at HERA very well.
As $z = E_{J/\psi}/E_\gamma$ goes to the endpoint, $z \to 1$, the NRQCD
prediction begins to rise.  This has been dubbed the ``HERA anomaly'',
and has been used to argue that NRQCD is incorrect.

However, it is now known that the NRQCD prediction breaks down near
the boundary of phase space, precisely where there is a problem at
HERA \cite{BRW,BSW}.  In this region the velocity expansion is
breaking down, and a shape function must be introduced.  In fact, one
should not compare the prediction with the data above $z\sim 1-v^2
\approx 0.7$.  If one looks only below this point, there is no problem
with the theory.  Unfortunately, it is not really possible to extract
the matrix elements from the data in this region.

\subsection{$B\to J/\psi$ DECAYS}
For $B\to J/\psi$ decays, the first worry is that the corrections to
factorization will be large since there is not a lot of energy in the
decay products.  There are also $\alpha_s$ corrections \cite{BMR} and
corrections in the Heavy Quark Effective Theory $1/m_b$ expansion
\cite{Ma}, the most important being the Fermi motion of the b-quark,
which requires the use of another shape function.

Nevertheless, it is possible to obtain the linear combination
$M_k^{J/\psi}$ from $B$ decays, with qualitative agreement with the
Tevatron extractions \cite{BMR,Ma}
\begin{eqnarray}
M_{3.1}^{J/\psi} = 1.5\times10^{-2}{\rm\ GeV^3},\\
M_{3.4}^{J/\psi} = 2.4\times10^{-2}{\rm\ GeV^3}.
\end{eqnarray}

\section{CONCLUSIONS}
NRQCD has been used for a number of years to make predictions about
quarkonium production.  At this time, there are consistent extractions
of the NRQCD matrix elements from a number of different experiments.
The most analysis has been done on hadronic collisions, in particular
at the Tevatron due to the high statistics.  

LEP is a very clean place to extract the matrix elements.  It is
possible obtain smaller theoretical uncertainty compared to the
Tevatron extractions.  Unfortunately the statistics are not as good.
Nevertheless, due to the large theoretical errors in the Tevatron
extractions, the LEP extraction is the best extraction to date.

It is also possible to test the NRQCD formalism at LEP by looking at
the moments of the spectrum.  Without the color octet channels
included in the NRQCD formalism, the theory and experiment would not
agree for this test.  This is one place where NRQCD passes a test.  It
will be interesting to see whether it will be possible to resolve the
polarization problem within the NRQCD factorization formalism.

I would like to thank my collaborators, Glenn Boyd, Peter Cho, and Ira
Rothstein, and the organizers of the IV International Conference on
Hyperon, Charm and Beauty Hadrons.  This work was supported in part by
the Department of Energy under grant number DOE-ER-40682-143.


\begin{thebibliography}{9}
\bibitem{BBL}
G.~T.~Bodwin, E.~Braaten and G.~P.~Lepage, Phys.\ Rev.\ {\bf D51},
(1995) 1125; erratum Phys.\ Rev.\ {\bf D55}, (1997) 5853.

\bibitem{recent}
For recent work on NRQCD, see M.~E.~Luke, A.~V.~Manohar and
I.~Z.~Rothstein, Phys.\ Rev.\ {\bf D61}, 074025 (2000), and 
references therein.

\bibitem{BF}
E.~Braaten and S.~Fleming, Phys.\ Rev.\ Lett.\  {\bf 74} (1995) 3327.

\bibitem{CW}
P.~Cho and M.~B.~Wise, Phys.\ Lett.\ {\bf B346} (1995) 129; M.~Beneke
and I.~Z.~Rothstein, Phys.\ Lett.\  {\bf B372} (1996) 157; M.~Beneke
and M.~Kramer, Phys.\ Rev.\ {\bf D55} (1997) 5269; A.~K.~Leibovich,
Phys.\ Rev.\ {\bf D56} (1997) 4412; E.~Braaten, B.~A.~Kniehl and
J.~Lee, hep-ph/9911436.

\bibitem{Affolder}
T.~Affolder {\it et al.} [CDF Collaboration], hep-ex/0004027.

\bibitem{Kramer}
M.~Kr\"amer, these proceedings.

\bibitem{DSL}
J.~L.~Domenech and M.~A.~Sanchis-Lozano, Phys.\ Lett.\ {\bf B476}
(2000) 65; E.~Braaten, S.~Fleming and A.~K.~Leibovich, hep-ph/0008091.

\bibitem{CL1}
P.~Cho and A.~K.~Leibovich, Phys.\ Rev.\ {\bf D53} (1996) 150.; 
Phys.\ Rev.\ {\bf D53} (1996) 6203.

\bibitem{BK}
M.~Beneke and M.~Kramer, Phys.\ Rev.\ {\bf D55} (1997) 5269.

\bibitem{CSL}
B.~Cano-Coloma and M.~A.~Sanchis-Lozano, Nucl.\ Phys.\ {\bf B508}
(1997) 753.

\bibitem{SL}
M.~A.~Sanchis-Lozano, hep-ph/9907497.

\bibitem{KK}
B.~A.~Kniehl and G.~Kramer, Eur Phys J C6 (1999) 493.

\bibitem{KS} 
J.~H.~K\"uhn and H.~Schneider, Phys.\ Rev.\ {\bf D24} (1981) 2996;
Z.\ Phys.\ {\bf C11} (1981) 263.

\bibitem{BCK}
V.~Barger, K.~Cheung and W.~Y.~Keung, Phys.\ Rev.\ {\bf D41} (1990)
1541; E.~Braaten, K.~Cheung and T.~C.~Yuan, Phys.\ Rev.\ {\bf D48} (1993)
4230.

\bibitem{CKY}
K.~Cheung, W.~Keung and T.~C.~Yuan, Phys.\ Rev.\ Lett.\ {\bf 76}
(1996) 877; P.~Cho, Phys.\ Lett.\ {\bf B368} (1996) 171.

\bibitem{rus}
Yu.~L Dohkshitzer et.~al., ``Basics of Pertrubative QCD'', Editions
Frontiers, Gif-sur-Yvette (1991); A.~Bassetto, M.~Ciafaloni and
G.~Marchesini, Phys.\ Rep.\ {\bf 100} (1983) 201.

\bibitem{BLR}
C.~G.~Boyd, A.~K.~Leibovich and I.~Z.~Rothstein, Phys.\ Rev.\ {\bf
D59} (1999) 054016.

\bibitem{ALEPH}
ALEPH Collaboration, submission to the 1997 EPS-HEP conference, No. 624.

\bibitem{BRW}
I.~Z.~Rothstein and M.~B.~Wise, Phys.\ Lett.\  {\bf B402} (1997) 346;
M.~Beneke, I.~Z.~Rothstein and M.~B.~Wise, Phys.\ Lett.\ {\bf B408}
(1997) 373.

\bibitem{BSW}
M.~Beneke, G.~A.~Schuler and S.~Wolf, hep-ph/0001062.; S.~Wolf, these
proceedings, hep-ph/0008180.

\bibitem{BMR}
M.~Beneke, F.~Maltoni and I.~Z.~Rothstein, Phys.\ Rev.\ {\bf D59}
(1999) 054003.  

\bibitem{Ma}
J.~P.~Ma, hep-ph/0006060.

\end{thebibliography}
\end{document}